\documentclass[twocolumn,aps,pra,superscriptaddress,showpacs]{revtex4-2}
\usepackage{epsfig}
\usepackage[colorlinks=true,citecolor=blue,urlcolor=blue,linkcolor=blue]{hyperref}
\usepackage{journal}
\usepackage{color} 
\usepackage[table,xcdraw]{xcolor}

\usepackage{tabularray}
\UseTblrLibrary{booktabs, multirow, siunitx}

\definecolor{D8D6C1}{HTML}{D8D6C1}
\definecolor{ECEADF}{HTML}{ECEADF}

\usepackage{tabularx,booktabs}
\newcolumntype{Y}{>{\centering\arraybackslash}X}
\newcolumntype{Z}{D..{1.7}}
\newcolumntype{A}{D..{2.9}}
\definecolor{ms}{RGB}{0, 150, 0}

\usepackage{dcolumn}
\newcolumntype{d}[1]{D{.}{.}{#1}}
\usepackage{multirow}
\usepackage{rotating}
\usepackage{float}

\usepackage{amssymb}
\usepackage{enumitem} 
\usepackage{amsthm}
\usepackage{amsmath}

\definecolor{darkgreen}{rgb}{0.0, 0.6, 0.0}

\usepackage{lipsum}
\definecolor{lipsumgray}{rgb}{0.8,0.8,0.8}

\begin{document}


\title{Cavity-enhanced spectroscopy in the deep cryogenic regime -\\ new hydrogen technologies for quantum sensing}


\author{K. Stankiewicz}
\affiliation{Institute of Physics, Faculty of Physics, Astronomy and Informatics, Nicolaus Copernicus University in Torun, Grudziadzka 5, 87-100 Torun, Poland}

\author{M. Makowski}
\affiliation{Institute of Physics, Faculty of Physics, Astronomy and Informatics, Nicolaus Copernicus University in Torun, Grudziadzka 5, 87-100 Torun, Poland}

\author{M. S{\l}owi\'nski}
\affiliation{Institute of Physics, Faculty of Physics, Astronomy and Informatics, Nicolaus Copernicus University in Torun, Grudziadzka 5, 87-100 Torun, Poland}

\author{K.L. So{\l}tys}
\affiliation{Institute of Physics, Faculty of Physics, Astronomy and Informatics, Nicolaus Copernicus University in Torun, Grudziadzka 5, 87-100 Torun, Poland}

\author{B. Bednarski}
\affiliation{Institute of Physics, Faculty of Physics, Astronomy and Informatics, Nicolaus Copernicus University in Torun, Grudziadzka 5, 87-100 Torun, Poland}

\author{H.~J\'o\'zwiak}
\affiliation{Institute of Physics, Faculty of Physics, Astronomy and Informatics, Nicolaus Copernicus University in Torun, Grudziadzka 5, 87-100 Torun, Poland}

\author{N. Stolarczyk}
\affiliation{Institute of Physics, Faculty of Physics, Astronomy and Informatics, Nicolaus Copernicus University in Torun, Grudziadzka 5, 87-100 Torun, Poland}

\author{M. Naro\.znik}
\affiliation{Institute of Physics, Faculty of Physics, Astronomy and Informatics, Nicolaus Copernicus University in Torun, Grudziadzka 5, 87-100 Torun, Poland}

\author{D. Kierski}
\affiliation{Institute of Physics, Faculty of Physics, Astronomy and Informatics, Nicolaus Copernicus University in Torun, Grudziadzka 5, 87-100 Torun, Poland}

\author{S. W\'ojtewicz}
\affiliation{Institute of Physics, Faculty of Physics, Astronomy and Informatics, Nicolaus Copernicus University in Torun, Grudziadzka 5, 87-100 Torun, Poland}

\author{A.~Cygan}
\affiliation{Institute of Physics, Faculty of Physics, Astronomy and Informatics, Nicolaus Copernicus University in Torun, Grudziadzka 5, 87-100 Torun, Poland}

\author{G. Kowzan}
\affiliation{Institute of Physics, Faculty of Physics, Astronomy and Informatics, Nicolaus Copernicus University in Torun, Grudziadzka 5, 87-100 Torun, Poland}

\author{P. Mas{\l}owski}
\affiliation{Institute of Physics, Faculty of Physics, Astronomy and Informatics, Nicolaus Copernicus University in Torun, Grudziadzka 5, 87-100 Torun, Poland}

\author{M. Piwi\'nski}
\affiliation{Institute of Physics, Faculty of Physics, Astronomy and Informatics, Nicolaus Copernicus University in Torun, Grudziadzka 5, 87-100 Torun, Poland}

\author{D. Lisak}
\affiliation{Institute of Physics, Faculty of Physics, Astronomy and Informatics, Nicolaus Copernicus University in Torun, Grudziadzka 5, 87-100 Torun, Poland}

\author{P. Wcis{\l}o}
\email[]{piotr.wcislo@fizyka.umk.pl}
\affiliation{Institute of Physics, Faculty of Physics, Astronomy and Informatics, Nicolaus Copernicus University in Torun, Grudziadzka 5, 87-100 Torun, Poland}

\date{\today}

\begin{abstract}

Spectrometers based on high-finesse optical cavities have proven to be powerful tools for applied and fundamental studies. Extending this technology to the deep cryogenic regime is beneficial in many ways: Doppler broadening is reduced, peak absorption is enhanced, the Boltzmann distribution of rotational states is narrowed, all unwanted molecular species disturbing the spectra are frozen out, and dense spectra of complex polyatomic molecules become easier to assign. We demonstrate a cavity-enhanced spectrometer fully operating in the deep cryogenic regime down to 4~K. We solved several technological challenges that allowed us to uniformly cool not only the sample but also the entire cavity, including the mirrors and cavity length actuator, which ensures the thermodynamic equilibrium of a gas sample. Our technology well isolates the cavity from external noise and cryocooler vibrations. This instrument enables a variety of fundamental and practical applications. We demonstrate a few examples based on accurate spectroscopy of cryogenic hydrogen molecules: accurate test of the quantum electrodynamics for molecules; realization of the primary SI standards for temperature, concentration and pressure in the deep cryogenic regime; measurement of the H$_2$ phase diagram; and determination of the ortho-para spin isomer conversion rate.

\end{abstract}



\maketitle

Cavity-enhanced spectroscopy has revolutionized various fields of research by providing unprecedented sensitivity and precision \cite{Gianfrani2024}. It has been instrumental in atmospheric studies, enabling the detection of pollutants and greenhouse gases \cite{Gomez-Pelaez2019} at parts-per-trillion levels and sub-permille accuracies \cite{Bielska2022}. This technique has enabled realization of standards for ultralow humidity \cite{Lisak2009} and carbon isotope ratios \cite{Fleisher2021}, and testing the quantum electrodynamics (QED) \cite{Zaborowski2020}. Even more sensitive noise-immune cavity-enhanced optical heterodyne molecular spectroscopy (NICE-OHMS) \cite{Ye1996} has become invaluable in fundamental physics \cite{Cozijn2023}. New developments have extended cavity-enhanced spectroscopy to study fast processes, large molecules and multispecies samples by combining it with optical frequency combs \cite{Bernhardt2010}. The accuracy was further improved with rapid scanning \cite{Truong2013} and purely frequency-based methods \cite{Cygan2024}.

Molecular spectroscopy benefits in many aspects from operating in the deep cryogenic regime, often far below 77~K. In this regime, Doppler broadening is reduced, peak absorption is enhanced, the Boltzmann distribution of rotational states is narrowed, all unwanted molecular species disturbing the spectra \cite{Fasci2018} are frozen out, and dense spectra of complex polyatomic molecules become easier to assign \cite{Changala2019}. Accurate spectroscopy at low temperatures is also important for testing state-of-the-art \textit{ab initio} quantum-chemical calculations, especially the interactions between atoms and molecules for both bound \cite{Jankowski2012} and scattering \cite{Slowinski2020} states. The quantum-scattering phenomena (collision-induced line-shape effects) are particularly interesting in the deep cryogenic regime (below 50~K), where fundamental discrepancies between experimental results and \textit{ab initio} calculations have been reported for many molecular species \cite{Thachuk1996} without a clue to resolve them. Cryogenic spectroscopy is also important for providing reference molecular data for astrophysical studies, particularly for studying the atmospheres of planets and exoplanets \cite{Jozwiak2024}. 

A typical approach to optical spectroscopy at cryogenic temperatures is based on Fourier spectrometers \cite{McKellar1990, Sung2023}. The Grenoble group developed cavity ring-down spectrometer (CRDS) with a gas sample kept at temperatures down to 77~K \cite{Kassi2022}. A phase-shift CRDS operating at 105~K was demonstrated in Ref.~\cite{Moehnke2006}. Optical cavities were also combined with buffer-gas-cooled molecular samples  \cite{Santamaria2016, Changala2019, Libert2024} and molecular beams \cite{Didriche2010,Perot2024}. Recently, high-resolution cryogenic cavity-enhanced spectrometers were developed by the Hefei \cite{Liu2022} and Amsterdam \cite{Cozijn2023} groups. In the Hefei spectrometer, the cavity mirrors are kept at room temperature while cryogenic gas flows through the cell \cite{Liu2022}. In the Amsterdam spectrometer, the entire optical cavity is uniformly cooled to 72~K~\cite{Cozijn2023}. Cryogenic optical cavities (operating at $T=1.7$ to 10~K) are also used as state-of-the-art short-term frequency references \cite{Robinson2019,Wiens_2023,He_2023,Valencia_2024}.

\begin{figure*}
    \centering
    \includegraphics[width=\textwidth]{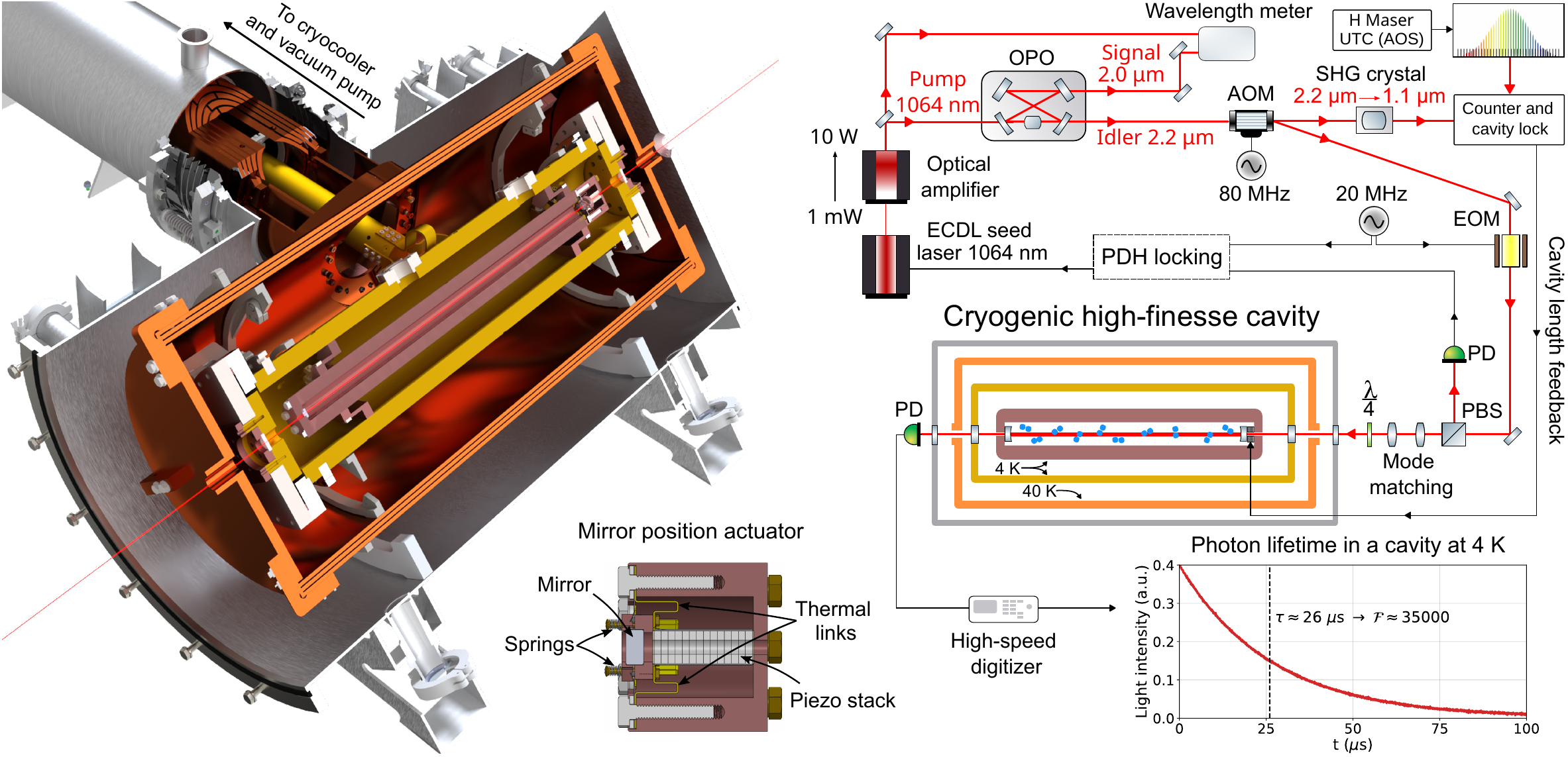}
    \caption{\textbf{Spectrometer based on a cryogenic high-finesse cavity.} \textbf{(left panel) Illustration of the cryogenic part of the setup.} The entire cavity, including the mirror position actuator, is placed inside a cryogenic copper vacuum chamber (yellow) that can be cooled to 4~K ensuring highly homogeneous thermodynamic equilibrium of the entire gas sample. The outer copper layers (orange), kept at approximately 40~K, block room-temperature thermal radiation. The entire system is placed inside a room-temperature vacuum system (gray) to maintain cryogenic conditions. The inner cryogenic and outer room-temperature vacuum chambers are independent; hence, the cryogenic chamber can be independently filled with H$_2$ gas. The inner cryogenic vacuum chamber is suspended on thin titanium rods (not shown here) to eliminate thermal links and attenuate vibrations. The cryogenic mirror position actuator \cite{Slowinski2022} is shown in detail in the small inset below the vacuum chamber. \textbf{(right panel)} \textbf{Simplified scheme of the experimental setup.} We use an optical parametric oscillator (OPO) to access the H$_2$ fundamental-band transition with an idler beam operating at 2.2 $\mu$m. The idler beam is tightly locked to the cryogenic high-finesse cavity via a PDH lock. The absolute frequency after idler frequency doubling is determined by referring to an optical frequency comb at 1.1 $\mu$m. The length of the cryogenic cavity is actively stabilized by referring to the optical frequency comb. Both the comb repetition rate and the offset frequency are stabilized to the radio frequency from the hydrogen maser (H maser), which is integrated with the local UTC (AOS).}
    \label{fig:expSetup}
\end{figure*}

In this work, we demonstrate the first cavity-enhanced spectrometer fully operating in the deep cryogenic regime down to 4~K. Not only the sample but also the entire cavity, including the mirrors and cavity length actuator, is uniformly cooled, ensuring thermodynamic equilibrium of a gas sample. The cavity is well isolated from external noise and cryocooler vibrations. This instrument enables a variety of fundamental and practical applications. We demonstrate a few examples based on accurate spectroscopy of cryogenic hydrogen molecules: accurate test of the QED for molecules, optical realization of three SI units in the deep cryogenic regime, measurement of H$_2$ phase diagram, and measurement of ortho-para spin isomer conversion rate.

\section{Results}

The cryogenic spectrometer is shown in Fig.~\ref{fig:expSetup}. The greatest advantage and main novelty of this spectrometer is that it allows nearly perfect thermalization of the entire gas sample down to 4~K. We uniformly cool not only the H$_2$ molecules but also the entire inner cryogenic vacuum chamber, cavity spacer, mirrors, and mirror position actuator. This feature is critical for accurate H$_2$-based optical metrology as temperature nonuniformity distorts the shape of an optical molecular resonance and hence introduces systematic uncertainties. The second important feature of this setup is the system of heavy copper blocks and layers suspended on thin titanium rods and thermally linked by flexible copper connectors that together make a mechanical system that well attenuates both internal (from the cryocooler and vacuum pump) and external vibrations and noise.

One of the cavity mirrors is installed on a position actuator (see the inset under the vacuum chamber in Fig.~\ref{fig:expSetup}) that allows the cavity length to be tuned by a few wavelengths. The actuator is designed to operate over a wide temperature range from room temperature down to 4~K~\cite{Slowinski2022}. A key feature of the actuator design is that it does not become misaligned when the temperature is cycled between 4 and 300~K (i.e., we initially align the optical cavity at room temperature and it remains aligned after the temperature is tuned within this range). The cavity has a length of 69~cm (FSR of 217 MHz) and finesse of 35~000 (photon lifetime of 26~$\mu$s). It is composed of two dielectric concave mirrors with a radius of curvature of 1~m. The length of the cavity is actively stabilized by locking frequency of one of the cavity resonances to optical frequency comb; see the cavity-length feedback line in Fig.~\ref{fig:expSetup}.

\begin{figure*}
    \centering
    \includegraphics[width=\textwidth]{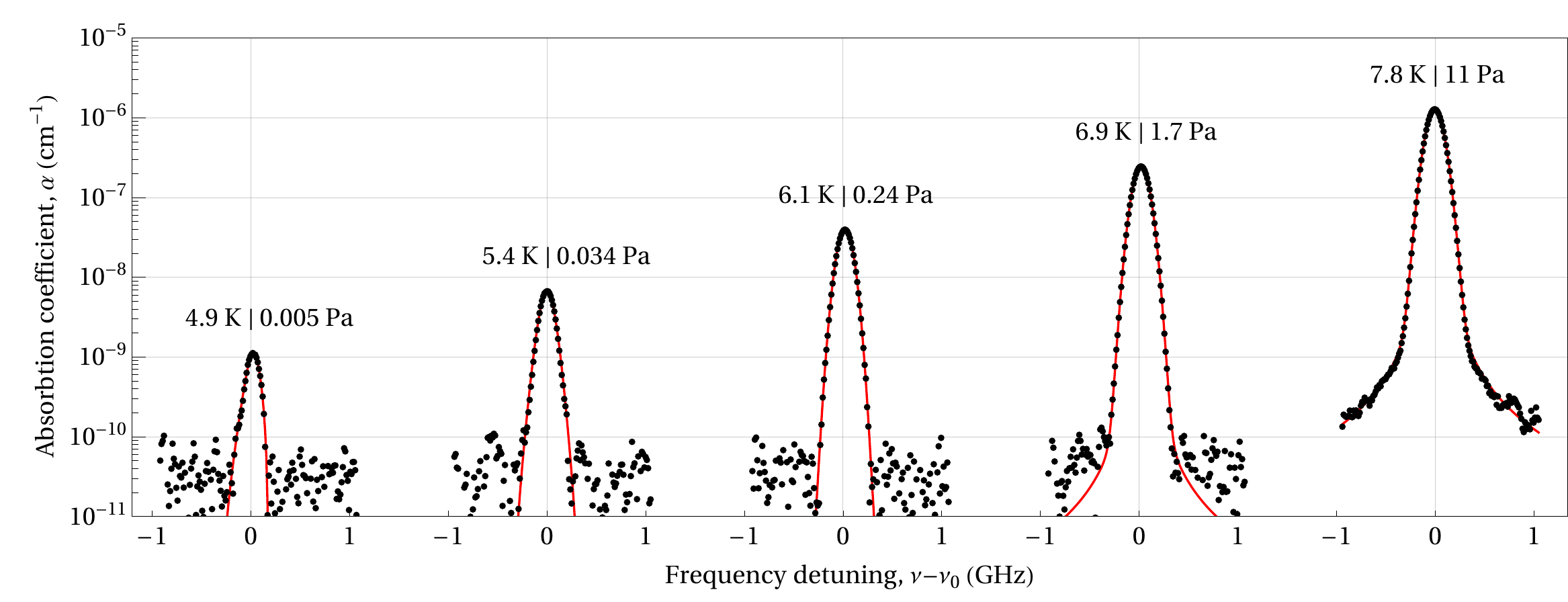}
    \caption{\textbf{Example of the spectra recorded with our instrument in the deep cryogenic regime.} The 1-0 S(0) line of the H$_2$ molecule at 2.22~$\mu$m is recorded at several temperatures from 4.9 to 7.8~K, which are far below the H$_2$ melting and boiling points where the vapor pressure strongly depends on temperature. The horizontal axis is shifted to the unperturbed position of the transition, $\nu_0$. The red line is the fitted line-shape model that includes both Doppler broadening and relevant collisional effects. Note the log scale on vertical axis.  }
    \label{fig:FitExamplesLogLin}
\end{figure*}

To fully demonstrate the capabilities of this experimental platform, we use a hydrogen molecule, which is the molecule with the highest equilibrium vapor pressure at deep cryogenic temperatures. To access the strongest band of H$_2$ (i.e., the fundamental band at 2.2 - 2.4~$\mu$m), we developed a laser system (similar to that recently reported in Ref.~\cite{Schenkel_2024}) based on a continuous-wave optical parametric oscillator (CW OPO); see the right panel in Fig. 1. A spectrally narrow (10~kHz) CW seed laser operating at 1064~nm after amplification to 10~W pumps an OPO cavity. In this work, we focus on the 1-0 S(0) transition at 2.22~$\mu$m, which is covered by the OPO idler beam tuning range. In contrast to the signal beam whose frequency is fixed by the OPO cavity, the idler beam inherits the seed laser frequency modulation. This allows us to tightly lock the idler beam to the high-finesse cryogenic cavity using the Pound-Drever-Hall (PDH) technique with the PDH feedback signal directly sent to the seed laser, see Fig.~\ref{fig:expSetup}. The absolute frequency of the idler beam is determined by referring to the optical frequency comb after doubling the idler frequency in a second-harmonic-generation (SHG) crystal. This also allows stabilization of the cavity length.

In the bottom right corner of Fig.~\ref{fig:expSetup}, we show the cavity photon lifetime recorded at 4~K. The high finesse is preserved when the temperature of the entire cavity is cycled between 4~K and room temperature. Molecular hydrogen ice is deposited at deep cryogenic temperatures on all inner surfaces (including mirrors); however, we do not observe deterioration of the mirror reflectivity (the finesse remains the same). Figure~\ref{fig:FitExamplesLogLin} shows an example of the cavity-ringdown spectra recorded with our instrument, the 1-0 S(0) line of the H$_2$ molecule at 2.22~$\mu$m. The wide dynamic range of the absorption coefficient, i.e., four orders of magnitude has already been demonstrated, with much room for further improvement, allows us to record a clear spectrum even below 5~K, far below the H$_2$ boiling and melting points. To our knowledge, this is the lowest temperature at which high-resolution cavity-enhanced spectroscopy at thermal equilibrium has been performed. In Sections~\ref{sec:application1}-\ref{sec:application4}, we present several examples of the applications of this instrument that address scientific and technological challenges in different fields.

\begin{figure*}
    \centering
    \includegraphics[width=\textwidth]{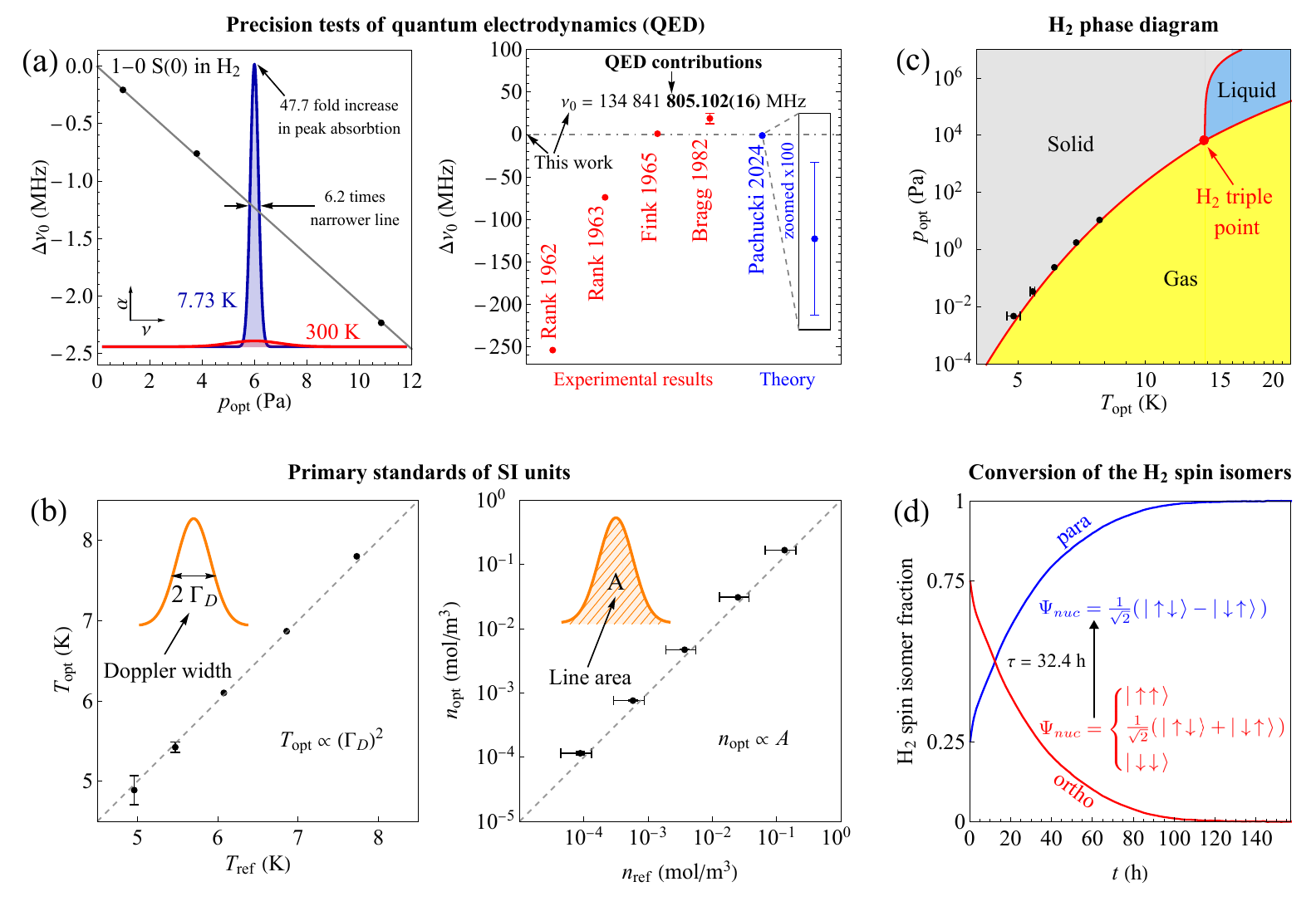}
    \caption{ \textbf{Examples of demonstrated applications of the cryogenic cavity-enhanced spectrometer. (a)~Accurate test of quantum electrodynamics (QED) for molecules.} Left panel: Position of the 1-0 S(0) line of H$_2$ as a function of pressure (black points) and its extrapolation to the zero-pressure limit (gray line). The red and blue lines are the simulations of the line at the same H$_2$ concentration but different temperatures, illustrating the metrological gain due to operation in the deep cryogenic regime. Right panel: Difference, $\Delta\nu_0$, between $\nu_0$ determined in this work along with previous experimental results \cite{Rank1962a, Rank1962b,Rank1963,Fink1965,Bragg1982} and the most recent theoretical value \cite{Pachucki2024}. \textbf{(b) Optical realization of SI units in the deep cryogenic regime.} $T_{\textrm{opt}}$ and $n_{\textrm{opt}}$ are the optically determined temperature and amount of substance per volume, respectively, whereas $T_{\textrm{ref}}$ and $n_{\textrm{ref}}$ are the same quantities simultaneously measured with independent techniques. \textbf{(c) Measurement of the H$_2$ phase diagram.} The black points are the results obtained in this work with the fully optical approach. The red lines are from Ref.~\cite{Souers1986}. \textbf{(d)~Ortho-para H$_2$ spin isomer conversion on copper surfaces.} The blue line comes directly from our measurement, whereas the red line is determined based on the condition that the total number of molecules inside the cavity is fixed. The lack of error bars in all the panels means that the uncertainties are too small to be visible at this scale, except for the first three points in the right plot in panel (a), for which the uncertainties are large but are not reported in Refs.~\cite{Rank1962a,Rank1962b, Rank1963,Fink1965}.} 
    \label{fig:Sbranch}
\end{figure*}

\subsection{Precision tests of quantum electrodynamics (QED) for molecules}
\label{sec:application1}

Accurate spectroscopy of simple atomic and molecular systems, whose structure can be calculated from the first principles of quantum theory, allows not only testing the QED sector of the standard model at high precision \cite{Zaborowski2020} but also searching for new physics beyond the standard model \cite{Salumbides2013}. An important example of such systems is molecular hydrogen for which the accuracy of \textit{ab initio} calculations of the energies of the rovibrational transitions reaches 10 significant digits \cite{Komasa2019}, and the present accuracy of the leading (nonrelativistic) contribution reaching 13 significant digits \cite{Pachucki2018} paves the way for further significant progress.

In this section, we present complementary experimental results for the case of the 1-0 S(0) transition in the H$_2$ molecule. The experiment was carried out at $T=7.8$~K at which the vapor pressure is high enough to reach high signal-to-noise ratio. We benefit from operating under deep cryogenic conditions in several ways. Compared to room temperature, at $T=7.8$~K the Doppler broadening is 6.2 times smaller and the peak absorption is $47.7=6.2 \times 7.7$ times larger, with the 6.2 factor being due to the smaller Doppler broadening and the 7.7 factor being due to the collapse of the entire Boltzmann distribution to $j=0$; compare the blue and red curves in the left panel in Fig.~\ref{fig:Sbranch}~(a). Overall, the higher and narrower resonance potentially increases the accuracy of the line position determination at $T=7.8$~K by a factor of nearly 300 compared to room temperature. Furthermore, at low temperatures all unwanted molecular species with strong dipole transitions disturbing the spectrum \cite{Fasci2018} are frozen out.

We acquired the spectra at three pressures, 0.98, 3.8 and 10.8~Pa, and extrapolated the pressure-perturbed line position to the zero-pressure limit; see the black points and gray line in the left panel in Fig.~\ref{fig:Sbranch} (a). The details of the experimental conditions, spectra analysis and  uncertainty budget are given in Methods section. Our ultimate determination of the 1-0 S(0) transition frequency is
$$\nu_0=134~841~805.102(15)_{\textrm{stat}}(6)_{\textrm{sys}}\textrm{~MHz}.$$
This value already includes the recoil shift. The measured photons are blueshifted; hence, we subtracted the recoil shift, which is 20.02 kHz in this case, from the measured zero-pressure value. We improve the previous best measurement of this line \cite{Bragg1982} by three orders of magnitude; see the right panel in Fig.~\ref{fig:Sbranch} (a). The most recent theoretical value \cite{H2SPECTRE,Pachucki2024} is 
$$\nu_0=134~841~803.8(9)\textrm{~MHz}.$$
The experimental accuracy reached in this work corresponds to testing the relativistic and QED corrections to $\nu_0$ to the fifth significant digit \cite{Komasa2019}.

Our accuracy of $\nu_0$ is higher comparing to the previous most accurate Doppler-limited measurements of homonuclear isotopologues of molecular hydrogen \cite{Zaborowski2020,Lamperti2023,Fleurbaey2023} and that for the most accurate Doppler-limited measurements of HD isotopologue \cite{Castrillo2021, Kassi2022} whose dipole line intensities are much larger comparing to the quadrupole line considered here.

\subsection{Realization of the primary SI units at deep cryogenic temperatures}
\label{sec:application2}

The redefinition of the SI unit system in 2019 \cite{Machin2018} revolutionized metrology at the fundamental level. Instead of referring to unit artifacts, the SI units are defined based on several recently fixed constants (such as Boltzmann and Planck constants), and the only remaining measurable reference is the hyperfine transition frequency in cesium atoms. This opens exciting possibilities for accurate metrology and open-access dissemination of SI unit primary standards but also poses new scientific and technological challenges regarding the actual physical realization of redefined units, especially under difficult-to-access conditions such as the deep cryogenic regime. We address this problem for the case of the SI units of temperature (kelvin), amount of substance per unit volume (mole $\cdot$ meter$^{-3}$), and pressure (pascal). For the first time, we demonstrate that optical methods can be used to realize these SI units in the difficult-to-access 4 - 20~K temperature regime.  

The thermodynamic temperature, $T$, can be spectroscopically determined by measuring the Doppler component, $\Gamma_D$, of the broadening of a molecular line \cite{Daussy2007}
\begin{equation}
\label{eq:kelvinUnit}
    T=\frac{1}{2\ln 2}\left(\frac{c^2}{k_B}\right)\left(\frac{m_{H_2}}{\nu_0^2}\right)\Gamma_D^2.
\end{equation}
The first parentheses contain fixed constants defining the SI units. The second parentheses contain quantities determined experimentally, but their combined relative uncertainty is less than $10^{-9}$; hence, they can also be treated as an exact number. The only quantity that needs to be measured is the frequency span, $\Gamma_D$, which we do using the primary time standard. This method provides the kelvin unit based only on the primary time standard and physical constants; hence, it can be regarded as a primary temperature standard \cite{Machin2018} (it does not require any calibration to other temperature standards).

We demonstrate optical realization of the kelvin unit at $T=4.9$ to $7.8$~K based on the spectra shown in Fig.~2. In the left panel of Fig.~3 (b), we show the optically determined temperature, $T_{\textrm{opt}}$, against a reference temperature, $T_{\textrm{ref}}$, measured with a commercial  diode sensor; see Table~\ref{table1Methdos} for exact values of $T_{\textrm{opt}}$ and $T_{\textrm{ref}}$ and their uncertainties. The uncertainty in $T_{\textrm{opt}}$ depends on temperature. For the two lowest temperatures, the H$_2$ vapor pressure decreases to a level at which it clearly limits the spectra signal-to-noise ratio; hence, the statistical contribution dominates the uncertainty in $T_{\textrm{opt}}$, which reaches 180~mK at the lowest temperature and decreases to below 20~mK at $T>6$~K. The optically determined temperature agrees with the reference diode-sensor measurements within the estimated total combined uncertainty, $\sigma$, for all the data points except the highest temperature at which the deviation is 4$\sigma$. We do not know whether the 4$\sigma$ discrepancy is caused by our optical method or the reference diode sensor.

The amount of substance per unit volume, $n$ (mol m$^{-3}$), is proportional to the line area, $A$ (which has a unit of m$^{-1}$s$^{-1}$; see the areas under the red curves in Fig.~2),
\begin{equation}
    \label{eq:moleUnit}
    n=\frac{15}{4\pi^4}\left(\frac{c^6}{N_A\hbar^4}\right)\left(\frac{\alpha^{3} m_e^4}{\nu_0^{3}}\right)\xi^{-1} A,
\end{equation}
where $N_A$, $\alpha$, and $m_e$ are the Avogadro constant, fine-structure constant, and electron mass, respectively. This relation has the same structure as Eq.~(\ref{eq:kelvinUnit}), i.e., it links the SI unit (mol m$^{-3}$) to spectroscopically determined quantity, $A$, which is based only on the time unit (the m$^{-1}$s$^{-1}$ unit is linked to the s$^{-2}$ unit via the fixed value of the speed of light). Similar to Eq.~(\ref{eq:kelvinUnit}), the first and second parentheses contain, respectively, the fixed constants defining the SI units and quantities determined experimentally with negligibly small uncertainties. Equation~(\ref{eq:moleUnit}) also includes a dimensionless factor, $\xi$, involving transition moment that comes from \textit{ab initio} calculations (since H$_2$ is the simplest neutral molecule, the corresponding $\xi$ factor can be calculated with high accuracy, $\xi=6.124(6)\times10^{-3}$~\cite{Roueff_2019}). Equation~(\ref{eq:moleUnit}) should also contain the Boltzmann distribution; however, the populations of H$_2$ rotational excited states are completely negligible. In the right panel of Fig.~\ref{fig:Sbranch} (b), we show the optically determined amount of gas per volume, $n_\textrm{opt}$, against its reference value, $n_\textrm{ref}$ (see Methods for $n_\textrm{ref}$). The optical method allows us to reach, in the difficult-to-access cryogenic regime, exceptionally small uncertainty of the (mol m$^{-3}$) unit realization far below $1\%$ which is two orders of magnitude better than the values obtained from the previous best $p(T)$ dependence in Ref. \cite{Souers1986}.

\subsection{Hydrogen phase diagram}
\label{sec:application3}

Simultaneous optical realization of the SI units of temperature and amount of substance per volume unit, $T_\textrm{opt}$ and $n_\textrm{opt}$, as described in the previous section, also automatically gives the SI unit of pressure via the equation of state of a gas, which in the relevant low-pressure regime is $p_\textrm{opt}=k_{B}N_A n_\textrm{opt} T_\textrm{opt}$. Note that $T_\textrm{opt}$ and $n_\textrm{opt}$ are multiplied only by the exact SI defining constants; hence, such optical realization of the pascal unit can be regarded as a primary SI standard (i.e., it does not require any calibration to other intermediate pressure standards). This allowed us, for the first time, to measure a part of the phase diagram of molecular hydrogen with purely optical method; see Fig.~\ref{fig:Sbranch}~(c). This first experimental demonstration already covers more than three orders of magnitude of pressure and considerably improves the accuracy of the previous best $p(T)$ curve \cite{Souers1986} in this range; see Methods. One of the highly promising applications that can be realized with this technology is accurate purely optical determination of H$_2$ triple point, which not only is an important anchor point for calibration of cryogenic temperature sensors but also has recently gained interest from the perspective of its first-principle determination~\cite{Deiters_2023}.

\subsection{Conversion between H$_2$ spin isomers}
\label{sec:application4}

Molecular hydrogen exists in two distinct isomeric states, referred to as para- and ortho-H$_2$, which differ by the symmetry of the nuclear wavefunction and are related to the total nuclear spin $I$ ($I=0$ and $1$ for para- and ortho-H$_2$, respectively). Although the energy of the interaction between nuclear spins is extremely small (far below 1 millikelvin), the Pauli exclusion principle, which implies the relationship between the symmetries of the nuclear and rotational wave functions, result in an even rotational quantum number, $j$, being allowed only for para-H$_2$ and an odd $j$ for ortho-H$_2$, resulting in a large energy difference between the para and ortho isomers (above 170 kelvin). On the one hand, this makes the ortho-to-para H$_2$ conversion a highly exothermic process (the released heat exceeds the evaporation heat). On the other hand, the ortho-to-para transition requires a nuclear spin flip making the conversion process inherently slow. These features have important technological consequences, especially for liquid H$_2$ energy storage technologies~\cite{Tzimas2003}, implying the necessity of developing efficient ortho-para catalysts. Research on ortho-para conversion is also important from a basic-physics perspective, as it enables studies of the magnetic properties and structural composition of solid surfaces~\cite{Ilisca1991}, processes on nonmagnetic surfaces of noble metals~\cite{Ilisca1992} or various physical properties of solid hydrogen~\cite{Driesen1984}. Moreover, ortho-para conversion is crucial for astrophysics~\cite{Lique2014} and \textit{ab initio} quantum calculations~\cite{Pachucki2008}.

In this work, we demonstrate that using the experimental platform shown in Fig.~\ref{fig:expSetup}, we can precisely track (with fully optical methods) the composition of the H$_2$ spin isomer mixture. We use the fact that the optical determination of the H$_2$ concentration, $n$, discussed in Sec.~\ref{sec:application2}, is spin isomer selective. The measurement starts by filling the experimental cryogenic chamber with a new portion of H$_2$ (taken directly from the room-temperature chamber), which immediately thermalizes with the cryogenic surfaces, but the rotational distribution remains separated between the para and ortho states preserving the normal ($1:3$ ratio) para-ortho distribution. This sets the initial conditions for our measurement, i.e., 0.25 and 0.75 fractions of H$_2$ occupy the $j=0$ (para-H$_2$) and $j=1$ (ortho-H$_2$) states, respectively; see the blue and red curves at $t=0$ in Fig.~\ref{fig:Sbranch}~(d). The blue curve comes directly from our measurement of spectral line area and the red curve comes from the condition that the total number of H$_2$ molecules in the cavity is fixed. The temperature at which the conversion was measured ($6.7$~K) is much lower than the energy (divided by $k_B$) of the $j=1$ state (170.5~K); hence, the remaining ortho population (after isomer thermalization) is negligible, and asymptomatically the para and ortho fractions converge to one and zero, respectively. The ortho-to-para conversion time measured in this work (the $1/e$ decay time) is 32.4(1.1)~h. The conversion time of approximately 50~h from the literature~\cite{Strzhemechny2002} suggests that the residual magnetic properties of the components used to construct the cryogenic vacuum chamber and the optical cavity slightly enhance the ortho-to-para conversion.

\section{Discussion}

We demonstrated, for the first time, a cavity-enhanced spectrometer operating in the deep cryogenic regime down to 4~K. This experimental platform enables various types of optical-metrology applications that were not previously possible. We demonstrated several examples: the most accurate Doppler-limited spectroscopy of molecular hydrogen, the first optical realization of three SI units in the deep cryogenic regime, optical measurement of the H$_2$ phase diagram resulting in a considerable improvement in the previous best $p(T)$ curve, and optical determination of the ortho-para H$_2$ spin isomer conversion rate. The instrument offers many directions for further development. One direction is implementation of the frequency-agile, rapid scanning spectroscopy (FARS) technique \cite{Truong2013}, which will improve the signal-to-noise ratio of the measured spectra and further reduce instrumental systematic errors \cite{Zaborowski2020}. Another direction is implementation of purely frequency-based spectroscopic methods \cite{Cygan2024}, which will further reduce instrumental systematic errors and extend the dynamic range to higher pressures, which will be particularly beneficial for accessing the H$_2$ triple point. We also plan to benefit from the high power of the laser (the 3~W idler beam), which, together with the high finesse of the cavity, makes saturation of the H$_2$ rovibrational lines and working in the Doppler-free regime feasible, where the cryogenic environment is beneficial (compared with room temperature) owing to the larger peak absorption, larger time of flight through the laser beam, and better control of the H$_2$ pressure down to below mPa, with all the other species perfectly frozen out. 

This experimental platform opens a way to address many other challenges in various fields, such as accurate measurements of weakly bound molecular complexes, spectroscopy of buffer-cooled complex molecules (whose spectra are difficult to resolve at elevated temperatures), acquisition of reference spectroscopic data for astrophysical applications or study of collisional effects at few-wavenumber collision energies.

\section{Acknowledgments}
We thank M. Drewsen and J. Glässel for discussions on cryogenic technologies. This research was funded by the European Union (Grant No 101075678, ERC-2022-STG, H2TRAP). Views and opinions expressed are, however, those of the author(s) only and do not necessarily reflect those of the European Union or the European Research Council Executive Agency. Neither the European Union nor the granting authority can be held responsible for them. This work is a part of the 23FUN04 COMOMET project that has received funding  from the European Partnership on Metrology, co-financed from the European Union’s Horizon Europe Research and Innovation Programme and by the Participating States. Funder ID 10.13039/100019599. We gratefully acknowledge Polish high-performance computing infrastructure PLGrid (HPC Center: ACK Cyfronet AGH) for providing computer facilities and support within computational grant no. PLG/2024/017252. Created using resources provided by Wroclaw Centre for Networking and Supercomputing (http://wcss.pl). The work was supported by the National Science Centre in Poland through Projects Nos. 2019/35/B/ST2/01118 (K.S., M.M., K.L.S., B.B., H.J., N.S., M.P., D.L., P.W.), 2022/46/E/ST2/00282 (M.S., D.K.), 2021/42/E/ST2/00152 (S.W.). The research was conducted using infrastructure from National Laboratory of Photonic and Quantum Technologies and is part of the program of the National Laboratory FAMO in Toruń, Poland.



\bibliography{bibliography}
\clearpage
\newpage
\section*{Methods}

\subsection*{Cryogenic cavity design and the laser system}

The large cryogenic vacuum chamber (with a length of 79~cm), whose temperature can be controlled down to 4~K, lies at the heart of the setup; see the thick yellow layer in Fig.~\ref{fig:expSetup}. The cryogenic vacuum chamber is independent of the outer room-temperature vacuum chamber (the gray layers in Fig.~\ref{fig:expSetup}), which allows us to fill the cryogenic chamber with a gas sample without introducing convection heat leaks between the cryogenic layers and the outer room-temperature chamber. The cryogenic high-finesse optical cavity (the thick brown cylinder with white mirrors at the ends in Fig.~\ref{fig:expSetup}) is installed inside the cryogenic vacuum chamber. Both the cavity spacer and the cryogenic vacuum chamber are made of high-purity copper and annealed after machining to ensure high thermal conductivity and, hence, excellent temperature uniformity. We use a low-vibration two-stage cryocooler \textit{PT420-RM} from \textit{Cryomech} with 1.8~W cooling power at 4.2~K.

The large masses of the cryogenic vacuum chamber and the thermal shields (130 and 200~kg, respectively) suspended on thin titanium and stainless steel rods, respectively, together with the high flexibility of the thermal connectors considerably attenuate the noise and vibrations. We do not observe any negative influence of the cryocooler vibrations on the optical cavity operation. Furthermore, we observe that the large mechanical inertia of the heavy components of the vacuum and cryogenic systems makes the optical cavity much more immune to external perturbations than typical room-temperature cavity-enhanced spectrometers.

In contrast to the previous design of the mirror position actuator reported in Ref.~\cite{Slowinski2022}, here, we do not glue the mirrors to the copper holders but use stainless steel springs that mechanically press the mirrors against the holders; see Fig.~\ref{fig:expSetup}. The thermal expansion mismatch between the copper mirror holder and the mirror substrate material leads to cavity misalignment at low temperature when the mirrors are glued, which is not the case when springs are used. The mirrors (from \textit{FiveNine Optics}) have dielectric coatings deposited on \textit{Corning 7979} (IR grade) fused silica substrates. The CW seed laser operating at 1064~nm and CW OPO are the \textit{Toptica CTL} and \textit{Toptica TOPO}, respectively.

\subsection*{Reference temperature and pressure determination}
The reference temperature, $T_{\textrm{ref}}$, was measured with a commercial diode sensor (the \textit{\mbox{DT-670}} diode in the \textit{\mbox{CU-HT}} package from \textit{Lake Shore Cryotronics, Inc.}, calibrated in the range from 1.4~K to 325~K) and read by cryogenic temperature monitor model \textit{218S} from the same manufacturer. $T_{\textrm{ref}}$ uncertainty varies from 20~mK at 4.9~K to 8~mK at 7.8~K, see Table~\ref{table1Methdos}.

The reference pressure, $p_{\textrm{ref}}$, was determined by combining $T_{\textrm{ref}}$ and the $p(T)$ curve taken from Ref.~\cite{Souers1986}
\begin{equation}
\label{eq:pT}
    p(T) = \exp\left( 9.2458    - 92.61    / T \right) T^{2.3794},
\end{equation} 
where $p$ and $T$ 
are given in Pa and K, respectively. This approach works in the vapor pressure regime, hence at 7.7~K in Table~\ref{table1Methdos} $p_{\textrm{ref}}$ is reported only for the highest pressure. The uncertainty of $p(T)$ for the relevant here temperature range is not reported in Ref.~\cite{Souers1986}.
Based on the uncertainties of the neighboring temperature ranges, we estimate $p_{\textrm{ref}}$ uncertainty at 50\%. Having the reference temperature and pressure, we determined the reference amount of substance per unit volume, $n_{\textrm{ref}}$, and its uncertainty using the equation
of state of a gas.



\subsection*{Spectra analysis}

In this work, we acquired the H$_2$ spectra at five temperatures from 4.9 to 7.8~K in the vapor pressure regime, which correspond to pressures from 4.68~mPa to 10.8~Pa, see Table~\ref{table1Methdos}. At the highest temperature, $T=7.8$~K, we acquired the spectra also at two lower pressures, 1 and 3.8~Pa, which we used to extrapolate the position of the 1-0 S(0) H$_2$ line to the zero-pressure limit. To increase the SNR, at each pressure-temperature point, we repeated the scans many times, see Table~\ref{table1Methdos}.

\begin{table*}[ht!]
    \centering
    \color{black}
    \caption{Optically determined temperature, $T_{\mathrm{opt}}$, pressure, $p_{\mathrm{opt}}$, and concentration $n_{\mathrm{opt}}$. $T_{\mathrm{ref}}$, $n_{\mathrm{ref}}$, and $p_{\mathrm{ref}}$ are the corresponding reference values determined with independent methods. $N_{\mathrm{scan}}$ is the number of scans of the H$_2$ 2-0 S(1) line acquired at each measurement point. At the highest temperature the spectra were collected at three pressures to have ability to extrapolate the line position to the zero-pressure regime. We determine $T_{\mathrm{opt}}$ at the lowest of these three pressures, because in this case the systematic uncertainty of $T_{\mathrm{opt}}$ related to pressure-induced line-shape effects is the smallest. The reference concentration, $n_{\mathrm{ref}}$, is available only at the highest of these three pressures at which the system is in the equilibrium vapor pressure regime. The numbers in parenthesis are the 1$\sigma$ standard combined uncertainties containing both statistical and systematic contributions.}
    \label{table1Methdos}
    \begin{tabularx}{0.9\linewidth}{lYYYYYYY}     \toprule
    $T_{\mathrm{opt}}$ (K)                  & $4.89(18)$    & $5.424(65)$   & $6.101(12)$  & $6.869(18)$ & -             & -           & $7.802(15)$   \\
    $T_{\mathrm{ref}}$ (K)                  & $4.96(2)$     & $5.471(9)$    & $6.076(8)$   & $6.858(8)$  & $7.735(8)$    & $7.733(8)$  & $7.731(8)$    \\\midrule
    $p_{\mathrm{opt}}$ (Pa)                 & $0.00467(37)$ & $0.03427(61)$ & $0.2399(11)$ & $1.759(9)$  & $10.851(42)$  & $3.808(15)$ & $0.9790(35)$  \\
    $p_{\mathrm{ref}}$ (Pa)                 & $0.0036(19)$  & $0.026(14)$   & $0.187(91)$  & $1.43(69)$  & $8.6(43)$     & -           & -             \\\midrule
    $n_{\mathrm{opt}}$ (10$^{-3}$mol/m$^3$) & $0.115(7)$    & $0.76(1)$     & $4.708(19)$  & $30.80(13)$ & $167.28(55)$  & $58.7(2)$   & $15.091(45)$  \\
    $n_{\mathrm{ref}}$ (10$^{-3}$mol/m$^3$) & $0.088(44)$   & $0.58(29)$    & $3.7(18)$    & $25(12)$    & $133(66)$     & -           & -             \\\midrule
    $N_{\mathrm{scan}}$                     & $12$          & $24$          & $10$         & $18$        & $14$          & $18$        & $20$          \\\bottomrule
    \end{tabularx}
\end{table*}

To analyzed the acquired spectra, we used the modified Hartmann-Tran (mHT) line-shape model \cite{Jozwiak2024} that includes the Doppler broadening and all the relevant collisional effects, i.e., the collisional broadening and shift, their speed dependencies, and the velocity-changing collisions with a complex Dicke parameter. The black points in Fig.~3~(a) demonstrate the dependence of the line position on pressure. They were determined by performing a single-pressure fit of the mHT profile to experimental spectra. The gray line is a linear extrapolation to the zero-pressure limit, $\nu_0$. It turns out, however, that more accurate determination of $\nu_0$ can be done when all the spectra at all the pressures are fitted simultaneously with a constraint that the collision-induced shift and line-shape parameters are linear in pressure. Furthermore, to reduce the systematic uncertainties originating from numerical correlations between $\nu_0$, line-shift, and line-shape parameters, we fixed the line-shape parameters to their theoretical \textit{ab initio} values, see Ref.~\cite{Cygan2024}. Following Ref.~\cite{Zaborowski2020}, to estimate the systematic uncertainties related to the description of the shape of the measured line, we repeated all the fits changing the values of the \textit{ab initio} line-shape parameters by a conservative amount of 10\%. In Table~\ref{Tab:UncertaintyBudget}, we show the uncertainty budget 
to our determination of the frequency of the 1-0 S(0) transition in H$_2$.

\begin{table}[h]
    \centering
    \caption{Standard uncertainty budget to experimental determination of the frequency of the 1-0 S(0) transition in H$_2$.}
    \label{Tab:UncertaintyBudget}
    \begin{tabularx}{0.45\textwidth}{lY}     \toprule
        \textbf{Uncertainty Contribution}      & $\mathbf{u(\nu_0)}$ \textbf{(kHz)} \\\midrule
        1. Line shape profile                  & $6$ \\
        2. Statistics, 1$\sigma$               & $15$\\
        3. Etalons                             & $<1$\\
        4. Temperature instability             & $<1$\\
        5. Relativistic asymmetry              & $<1$\\
        6. Laser source stability              & $<1$\\
        7. Recoil shift                        & $<1$ \\\midrule
        \textbf{Standard combined uncertainty} & $\mathbf{16}$\\\bottomrule
    \end{tabularx}
\end{table}

\subsection*{The amount of substance per unit volume}

Equation~(\ref{eq:moleUnit}) is valid for the quadrupole electric transitions, which is the case of the considered in this work 1-0 S(0) transition. In this case, the theoretical dimensionless $\xi$ factor is given by $\xi=(|Q_{\mathrm{eg}}|/ea_{0}^{2})^{2}$. Here, $Q_{\mathrm{eg}}$ denotes the quadrupole moment function of the ground electronic $^{1}\Sigma_{g}^{+}$ state of H$_{2}$, $Q(r)$~\cite{Roueff_2019}, averaged over the wavefunctions of the ground, $\chi_{\mathrm{g}}(r)$, and excited, $\chi_{\mathrm{e}}(r)$, rovibrational state of H$_{2}$
\begin{equation}
\label{eq:rovib_average_q_mom}
    Q_{\mathrm{eg}} = \int\mathrm{d}r \chi_{\mathrm{g}}(r) Q(r) \chi_{\mathrm{e}}(r),
\end{equation}
where $r$ is the internuclear distance in H$_{2}$, and $e$ and $a_{0}$ denote the elementary charge and Bohr radius, respectively.

In the case of the electric dipole lines (important for the HD isotopologue) Eq.~(\ref{eq:moleUnit}) takes the form
\begin{equation}
    \label{eq:moleUnit:E1}
    n=\frac{3}{4\pi^2}\left(\frac{c^2}{N_A\hbar^2}\right)\left(\frac{\alpha m_e^2}{\nu_0}\right)\xi^{-1} A,
\end{equation}
with the dimensionless theoretical factor ${\xi=(|d_{\mathrm{eg}}| / {ea_{0}})^{2}}$. Here, $d_{\mathrm{eg}}$ denotes the dipole moment function of the ground electronic $^{1}\Sigma^{+}$ state of HD, $d(r)$, averaged over the wavefunctions of the ground and excited rovibrational state of HD, as in Eq.~\eqref{eq:rovib_average_q_mom}.\\

\subsection*{H$_2$ phase diagram}

In this work, we improve the $p(T)$ curve from Ref.~\cite{Souers1986} by using our optically measured $(p,T)$ points from the H$_2$ phase diagram, the black points from
Fig.~\ref{fig:Sbranch}~(c). We fitted the three numerical parameters from Eq.~(\ref{eq:pT}) to a combined set of the old $(p,T)$ points used in Ref.~\cite{Souers1986} and our new much more accurate $(p,T)$ points resulting in 
\begin{equation}
\label{eq:newpT}
     p(T) = \exp\left( 9.5830    - 94.27    / T \right) T^{2.0054}.
\end{equation}
This new $p(T)$ curve is considerably more accurate than the one from Ref.~\cite{Souers1986}, especially in the $4.9-7.7$~K temperature range, where the 50\% estimated accuracy of Eq.~(\ref{eq:pT}) is improved to below 1\% for Eq.~(\ref{eq:newpT}).

{\color{ms}



}

\end{document}